\def\selectedoptions{}
\ifx\empty\selectedoptions
  \def\selectedoptions{final}
\fi

\documentclass[
   \selectedoptions
  ]
  {aipproc}

\def\selectedlayoutstyle{6x9} 
\layoutstyle\selectedlayoutstyle

\SetInternalRegister\hbadness{8000} 

\newcommand\doingARLO[2][]{%
  \ifx\mmref\undefined #1\else #2\fi
}

\begin{document}

\title 
      []
      {\boldmath New CLEO Results for {$|V_{cb}|$} and $|V_{ub}|$}


\author{Roy A. Briere, representing the CLEO Collaboration}{
  address={Carnegie Mellon University; 5000 Forbes Ave.; Pittsburgh, PA 15213},
  email={rbriere@andrew.cmu.edu},
  thanks={}
}

\copyrightyear  {2001}

\begin{abstract}
We report recent measurements from CLEO of the first two moments 
of the photon energy spectrum for $b \to s\,\gamma$ decays 
and the hadronic recoil mass in $\bar{B} \to X_c \ell \bar{\nu}$.  
These physical quantities allow one to fix non-perturbative parameters 
occurring in calculations based on HQET and QCD.  
Predictions for semileptonic decay rates within this same framework 
depend in addition on the CKM matrix elements $V_{qq^\prime}$ 
governing quark mixing.  
We can thus extract $|V_{cb}|$ 
from the inclusive semileptonic decay rate of $B$ mesons, 
and $|V_{ub}|$ from the lepton endpoint spectrum of 
$\bar{B} \to X_u \ell \bar{\nu}$.  
Model dependence is reduced except for 
the assumption of quark-hadron duality.  
Finally, we update the classic measurement of 
$|V_{cb}|$ from $\bar{B} \to D^* \ell \bar{\nu}$ at zero recoil.  

\end{abstract}

\date{\today}

\maketitle

\section{Introduction}

Most of the ad-hoc parameters of the Standard Model are contained 
in the flavor sector.  
Mysteries of mixing, mass generation, and $CP$ violation all meet here.  
$B$ physics offers the possibility of direct measurements of two 
of the Cabibbo-Kobayashi-Maskawa (CKM) mixing-matrix elements, 
$|V_{cb}|$ and $|V_{ub}|$, as well as access to two others, 
$|V_{td}|$ and $|V_{ts}|$, via loop processes.  We concentrate here on 
determinations of the former pair via semileptonic $B$ decays.  
 
Experimentally, we measure the number of certain semileptonic decays; 
we can convert this to a branching ratio via knowledge of $N_{B\bar{B}}$ 
(the number of $B$ pairs present in our data sample) and finally 
to a partial width by using $B$ lifetime measurements from elsewhere.  
Both these partial widths and certain other kinematic `moments' may be 
calculated in a systematic Heavy Quark Effective Theory (HQET) 
\cite{Isgur:1989vq,Neubert:1994mb} and QCD expansion.  
The expressions depend on some a priori unknown non-perturbative parameters, 
most notably $\bar{\Lambda}, \lambda_1$ and $\lambda_2$.  
The kinematic moments will allow us to independently determine 
these parameters for use in a self-consistent way in other formulae.  

Expressions for semileptonic decay rates also depend on CKM matrix elements.  
One can thus extract $|V_{cb}|$ from the inclusive semileptonic rate, 
or perform a more intricate extraction of $|V_{ub}|$ from the rate at the 
lepton endpoint of $b \to u$ semileptonic decays.   
The partial width calculations rely on quark-hadron duality 
\cite{Bigi:2000rq}.  
That is, they assume that for sufficiently inclusive quantities, 
quark-gluon calculations can be used for observed hadronic processes.  
The chief issues are non-perturbative effects 
and the need to average over enough hadronic states, 
which is even more problematic for $|V_{ub}|$ from the endpoint.  
It is therefore desirable to compare these results to those obtained 
from previous methods; one such result is also updated here.  
This is our determination of $|V_{cb}|$ from $\bar{B} \to D^* \ell \bar{\nu}$ 
at zero-recoil as favored by more familiar HQET treatments.  

The CLEO II detector \cite{Kubota:1992ww} and the CLEO II.V upgrade with 
a silicon vertex detector \cite{Hill:1998ea} 
and new drift chamber gas \cite{Briere:1999} are described elsewhere.  
Overall, in CLEO II and II.V data, 
the $\Upsilon(4S)$ to continuum luminosity ratio is about 2.1 
and the effective $B\bar{B}$ cross section is 1.06 nb.  
The $b \to s\,\gamma$ and the $\bar{B} \to X_u \ell \bar{\nu}$ analyses 
both use the entire CLEO II and II.V datasets with a total luminosity 
of about 13.5 fb$^{-1}$ and about $9.7 \times 10^6$ $B\bar{B}$ pairs, 
while the $\bar{B} \to X_c \ell \bar{\nu}$ and 
the $\bar{B} \to D^* \ell \bar{\nu}$ analyses 
use only the CLEO II portion with about $3.3 \times 10^6$ $B\bar{B}$ pairs.  

\section{\boldmath The \MakeLowercase{$b \to s\,\gamma$} Photon Spectrum}

\begin{figure}
\caption{Left: Observed photon spectrum shown in a) with the 
         scaled continuum background prediction from off-resonance data, 
         and after continuum subtraction in b), where the 
         $B\bar{B}$ background is now displayed.  
         Right: Measured photon spectrum for $b \to s\,\gamma$ events.  
        }
\includegraphics[height=0.3\textheight]{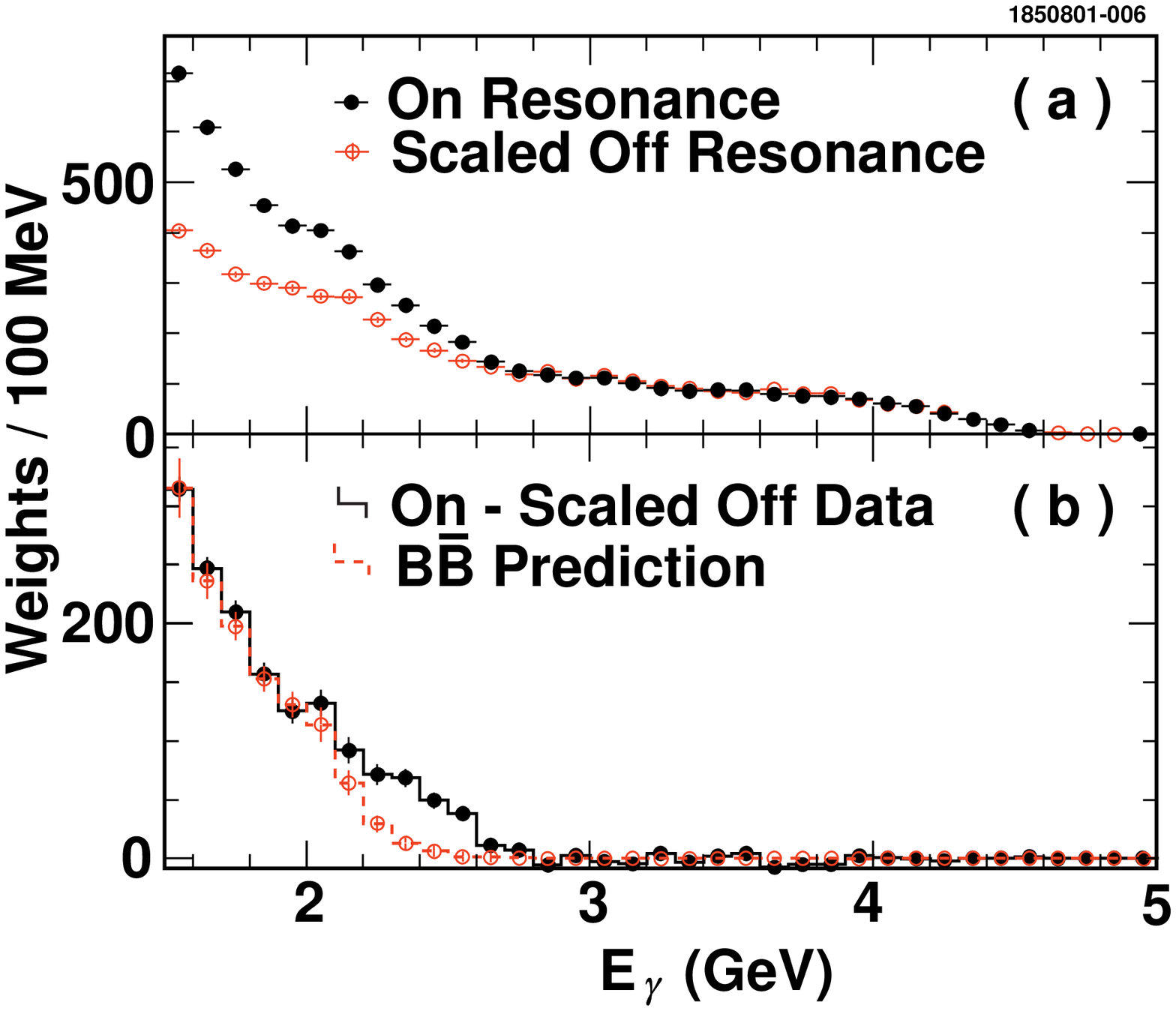}
\includegraphics[height=0.3\textheight]{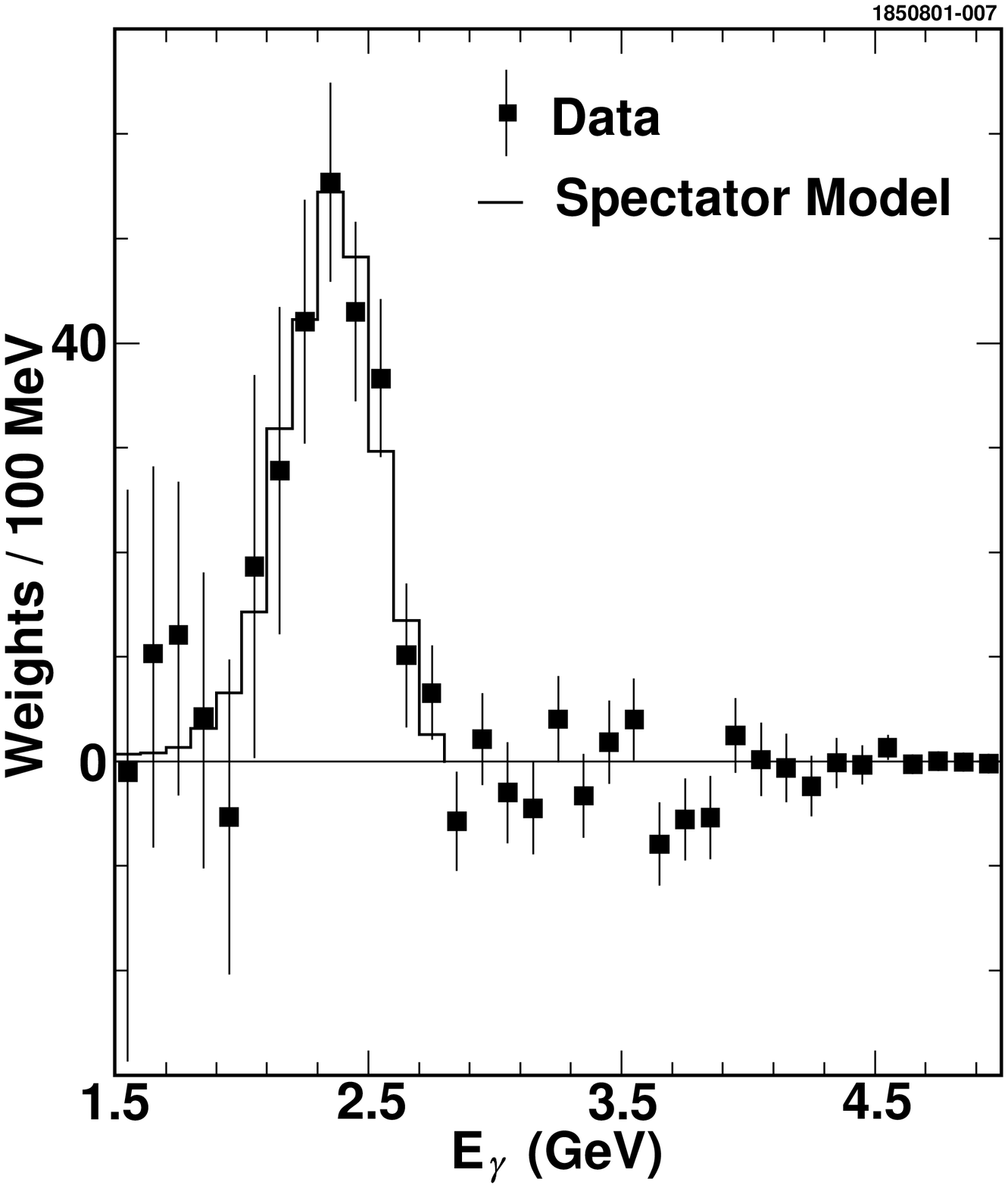}
\label{fig:photons}
\end{figure}

This final state is distinguished by a high-energy photon.  
Our new analysis uses a lower cut of 2.0 GeV on $E_\gamma$, 
reducing model-dependence systematics.  
Continuum background dominates, but a factor of one hundred suppression 
is achieved by using event shape information, kinematics of detected leptons, 
and a pseudo-reconstruction technique 
(seeking the best $K (n\pi)\, \gamma$ combination).  
A neural network combines information to give a signal weight 
for each event.  
Both the raw photon spectrum indicating background sources 
and the final subtracted spectrum are displayed 
in Figure \ref{fig:photons}.  
Though $B\bar{B}$ background increases at the lowest energies, 
we still see sensible behavior albeit with larger errors.  

Our  focus here is extracting moments of the $E_\gamma$ spectrum;   
for other analysis details, see \cite{Honscheid:hf}.  
This spectrum, naively a sharp line, has a width determined largely by the 
$b$ quark Fermi motion and somewhat by the varying recoil mass (i.e., QCD effects).  
Two smaller sources of width are the small known $B$ boost 
($\beta \simeq 0.06$) and resolution smearing.  

We determine moments directly from the data, accounting for 
energy-dependent efficiency, resolution, and boost smearing.  
As a check, we also take the  Ali-Greub \cite{Ali:1995bi} 
or Kagan-Neubert models \cite{Kagan:1998ym}, propagated through our 
full GEANT-based detector simulation, and take moments of these 
models with parameters which best fit the data.  
All extractions are consistent, and variations are reflected in 
the systematic error.  

We find, for $E_\gamma > 2.0$ GeV \cite{Chen:2001fj}: 
\begin{eqnarray}
               <E_\gamma>   &=& (2.346  \pm 0.032  \pm 0.011)  \,\,{\rm GeV}
\nonumber \\
<E_\gamma^2> - <E_\gamma>^2 &=& (0.0226 \pm 0.0066 \pm 0.0020) \,\,{\rm GeV}^2
\nonumber 
\end{eqnarray}
where the brackets $<...>$ denote the average value.  

The first moment, again with $E_\gamma > 2.0$ GeV, is calculated as 
\cite{Bauer:1998fe,Ligeti:1999ea,FalkLigeti:pc}: 
\begin{eqnarray}
\langle E_{\gamma}\rangle 
= \frac{M_B}{2} \,\,[\,\, 1 &-& 0.385 \frac{\alpha_s}{\pi} 
                     - 0.620 \beta_0 (\frac{\alpha_s}{\pi})^2 
\nonumber \\
&-& \frac{\bar \Lambda}{M_B} (1-.954 \frac{\alpha_s}{\pi} 
      - 1.175 \beta_0 (\frac{\alpha_s}{\pi})^2) 
\nonumber \\
&-& \frac{13 \rho_1 - 33 \rho_2}{12 M_B^3}  - \frac{{\cal T}_1 
   + 3{\cal T}_2 + {\cal T}_3 +3 {\cal T}_4}{4 M_B^3}
   - \frac{\rho_2 C_2}{9 M_D^2 M_B C_ 7}
\nonumber \\
&+& {\cal O}(1/M^4_B) \,\,]\,\,.
\nonumber
\end{eqnarray}

To lowest order, 
$\langle E_{\gamma}\rangle = \frac{1}{2} [M_B - \bar{\Lambda}]$.  
The parameter $\bar{\Lambda}$ measures 
the energy of the light degrees of freedom: the `brown muck' surrounding the 
heavy-quark.  
There are further parameters appearing in general at second order in $1/M_B$; 
they are absent in this particular case, and are discussed later.  
Finally, the third-order parameters 
${\cal T}_i$ and $\rho_i$ are estimated as ${\cal O}(0.5\, {\rm GeV}^3)$ 
with variations included as systematics.  

From the first moment alone, we find that 
$\bar{\Lambda} = (0.35 \pm 0.08 \pm 0.10)$ GeV \cite{Chen:2001fj}.  
Here, the entire experimental error is given first and the second error 
is due to the theoretical extraction.  
We avoid using second moments which give poorer determinations 
of the parameters and, in the case of hadronic moments below, 
have expressions which do not appear to converge as rapidly.
Since teh value of $\bar{\Lambda}$ is not meaningful out of context, 
one must do all calculations consistently with respect to scheme and order.  
We choose 
$\overline{MS}, \,\, {\cal O}(1/M_B^3), \,\, {\cal O}(\beta_0\alpha_s^2)$ 
everywhere.  

\section{Hadronic Moments in $\bar{B} \to X_\MakeLowercase{c} \ell \bar{\nu}$}

This analysis uses both $e$ and $\mu$ with $1.5 < p_\ell < 2.5$ GeV/$c$ and 
we reconstruct the neutrino properties via four-momentum balance 
with techniques developed at CLEO \cite{Alexander:1996qu}.  
Great care is taken to account exactly once for all observed particles; 
cuts on charge balance, a multiple-lepton veto, and 
$(E_{miss}^2 - p_{miss}^2)$ help ensure that the missing four-momenta 
is due to a single missing neutrino.  
The resolution on missing momentum is 
$\sigma(p_{miss}) \simeq 110\,{\rm  MeV}/c$.  
After checking consistency with zero missing mass, 
we use $E_\nu = |p_{miss}|$ rather than $E_{miss}$ since it has 
better resolution.  
We use $B$ decay kinematics to determine 
$M_X^2 = M_B^2 + M_{\ell\bar{\nu}}^2 - 2E_B E_{\ell\bar{\nu}} 
       + 2\vec{p}_B \cdot \vec{p}_{\ell\bar{\nu}}$ 
without needing to observe the hadrons or group them into those 
from the $B$ vs. the $\bar{B}$ decay.  
The small final dot product, which averages zero, is ignored.  
Thus, four-momentum balance of the entire $B\bar{B}$ event measures 
the neutrino properties, leaving the hadronic recoil system unobserved.  
A typical exclusive mode analysis would instead observe the 
hadron(s) and have an unobserved neutrino.  
We expect 95\% $b \to c \ell \bar{\nu}$ after continuum subtraction; 
the  rest ($c \to s \bar{\ell} \nu$ secondaries, $b \to u \ell \bar{\nu}$) 
is subtracted via Monte Carlo simulation.  

\begin{figure}
\caption{Left: Observed recoil mass in $\bar{B} \to X_c \ell \bar{\nu}$ 
         showing data as points along with a fit to $D$, $D^*$ and 
         heavier charm meson contributions.  
        Right: Constraints from measured $b \to s\,\gamma$ photon energy and 
        $\bar{B} \to X_c \ell \bar{\nu}$ recoil mass first moments 
        in the $\bar{\Lambda} - \lambda_1$ plane.  
        The ellipse indicates $\Delta \chi^2 = 1$, including 
        systematic errors.}
\includegraphics[height=0.3\textheight]{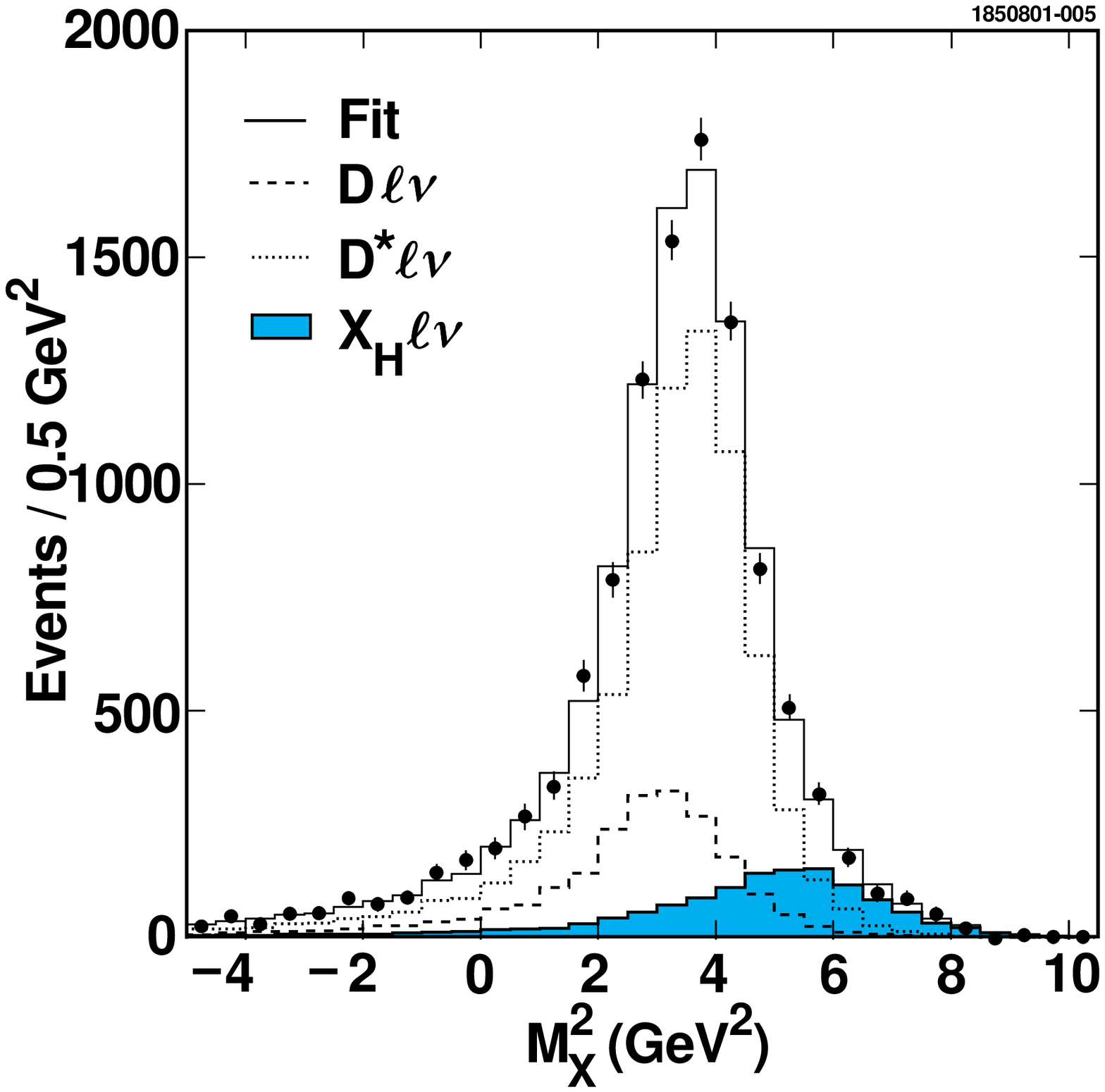}
\includegraphics[height=0.3\textheight]{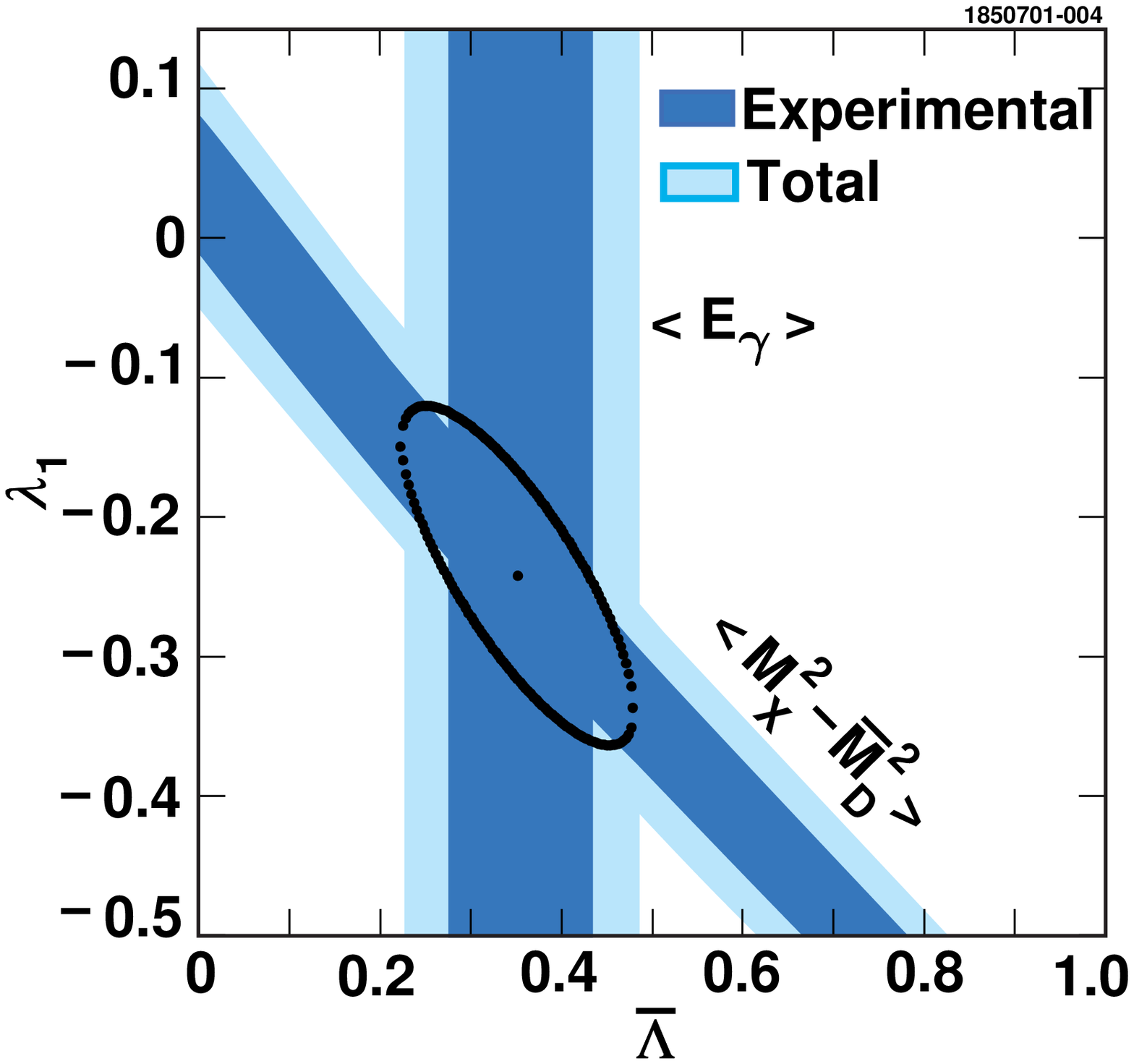}
\label{fig:mx2bands}
\end{figure}

Final results for the moments are calculated from 
the $M_X^2$ distributions corresponding to the mixture 
of $D \ell \bar{\nu}$, $D^* \ell \bar{\nu}$, and $X_H \ell \bar{\nu}$ 
spectra which best fit the data.  
We take moments of the generated $M_X^2$ distributions while fitting the 
data to reconstructed quantities passed through the full 
physics and detector simulation and 
analysis, hence accounting for the $B$ boost, resolution and efficiency.  
Heavy states $X_H$ beyond the $D$ and $D^*$ include $D^{**}$ states 
modeled with ISGW2 \cite{Isgur:1989gb}
and non-resonant $D^{(*)}\pi$ 
treated with the Goity-Roberts prescription \cite{Goity:1995xn}.  
Their normalization is fixed by data.  
The fit distributions are shown in the left panel 
of Figure \ref{fig:mx2bands}.  

We finally arrive at \cite{Cronin-Hennessy:2001fk}: 
\begin{eqnarray}
<M_X^2 - \bar{M}_D^2>     &=& (0.251 \pm 0.023 \pm 0.062) \,\,{\rm GeV}^2
\nonumber \\
<(M_X^2 - \bar{M}_D^2)^2> &=& (0.639 \pm 0.056 \pm 0.178) \,\,{\rm GeV}^4
\nonumber \\
<(M_X^2 - <M_X^2>)^2>     &=& (0.576 \pm 0.048 \pm 0.163) \,\,{\rm GeV}^4
\nonumber
\end{eqnarray}
where $\bar{M}_D$ denotes the spin-averaged $D, D^*$ mass.  
The main systematics include the neutrino reconstruction efficiency and 
the models of the $X_H$ states.  

The theoretical expressions for the 
moments \cite{Falk:1996kn,Gremm:1997df,Falk:1998jq} 
use a consistent scheme and include (most of) the effects 
of the lepton energy cut.  
Unlike the mean photon energy discussed earlier, 
the second order HQET expansion parameters appear here.  
These are $\lambda_1$, 
related to the Fermi motion energy of the $b$ quark, 
and $\lambda_2$, measuring the QCD hyperfine splitting; 
the latter is fixed from $m_{B^*} - m_B$ as measured by others.  
The third-order terms are treated as before.  

Combining with the $b \to s\,\gamma$ result, 
we find \cite{Cronin-Hennessy:2001fk}: 
\begin{eqnarray}
\bar{\Lambda} &=& (0.35  \pm 0.07  \pm 0.10)  \,\,{\rm GeV}
\nonumber \\
\lambda_1     &=& (-0.236 \pm 0.071 \pm 0.078) \,\,{\rm GeV}^2  
\nonumber 
\end{eqnarray}
The errors have the same meaning as for the $b \to s\,\gamma$ moments.  
The results are best viewed in the $\bar{\Lambda} -\lambda_1$ plane; 
see the right panel of Figure \ref{fig:mx2bands}.

\section{The Inclusive Semileptonic Rate and $|V_\MakeLowercase{cb}|$}

It is of course also possible to calculate the zeroth moment for 
semileptonic decays; 
this is simply $\Gamma_{sl} \equiv \Gamma(b \to c \ell\bar{\nu})$ 
\cite{Voloshin:1995cy,Shifman:1995jh,Ball:1995wa}.   
The expression looks like a free-quark decay, akin to the classic 
muon decay rate, with a phase-space factor for finite $m_c$, augmented by 
QCD corrections and the HQET expansion.  
The actual formula used, gleaned from calculations in 
\cite{Gremm:1997df,Bigi:1992su,Bigi:1993fe,Jezabek:1989iv,Luke:1995yc}, 
may be found in \cite{Cronin-Hennessy:2001fk}.  

Having consistently determined the lower-order HQET parameters
with moments, we are poised to extract $|V_{cb}|$.  
The inclusive semileptonic rate is taken from a venerable CLEO 
result \cite{Barish:1996cx} using the tagged di-lepton method 
\cite{Albrecht:1993pu}.  
After subtracting 1\% from the published value 
to correct for $\bar{B} \to X_u \ell \bar{\nu}$, we have:  
${\cal B}(\bar{B} \to X_c \ell \bar{\nu}) = (10.39 \pm 0.46)\%$.  

We convert to $\Gamma_{sl}$ using 
$\tau_{B^\pm} = (1.548 \pm 0.032)$ ps \cite{Groom:2000in}, 
$\tau_{B^0}   = (1.653 \pm 0.028)$ ps \cite{Groom:2000in}, 
and $f_{+-}/f_{00} = 1.04 \pm 0.08$ \cite{Alexander:2000tb}.  
Combining, we finally arrive at \cite{Cronin-Hennessy:2001fk}: 
\begin{eqnarray}
|V_{cb}| &=& (4.04 \pm 0.09 \pm 0.05 \pm 0.08) \times 10^{-2}   
\nonumber
\end{eqnarray}

The largest errors are from (in order) the 
measurement of $\Gamma_{sl}$, 
the HQET parameters $\bar{\Lambda}, \lambda_1$, 
and the scale for $\alpha_s$.  
This yields a precise ($3.2\%$) determination, but 
given the global quark-hadron duality issues 
one should compare to results from other methods.  

\section{\boldmath Extracting $|V_\MakeLowercase{ub}|$ from the 
$\bar{B} \to X_\MakeLowercase{u} \ell \bar{\nu}$ Endpoint}

\begin{figure}
\caption{Left: typical prediction of the lepton spectra for 
         $b \to c \ell \bar{\nu}$ and $b \to u \ell \bar{\nu}$ 
         (note the x10 here!).  
         Right: inclusive leptons in data near the endpoint.  
         Plot a) gives the raw spectrum and shows the continuum (shaded) 
         and $b \to c$ (open histogram) contributions; 
         plot b) shows the extracted efficiency-corrected 
         $b \to u \ell \bar{\nu}$ rate.  
        }
\includegraphics[height=0.3\textheight]{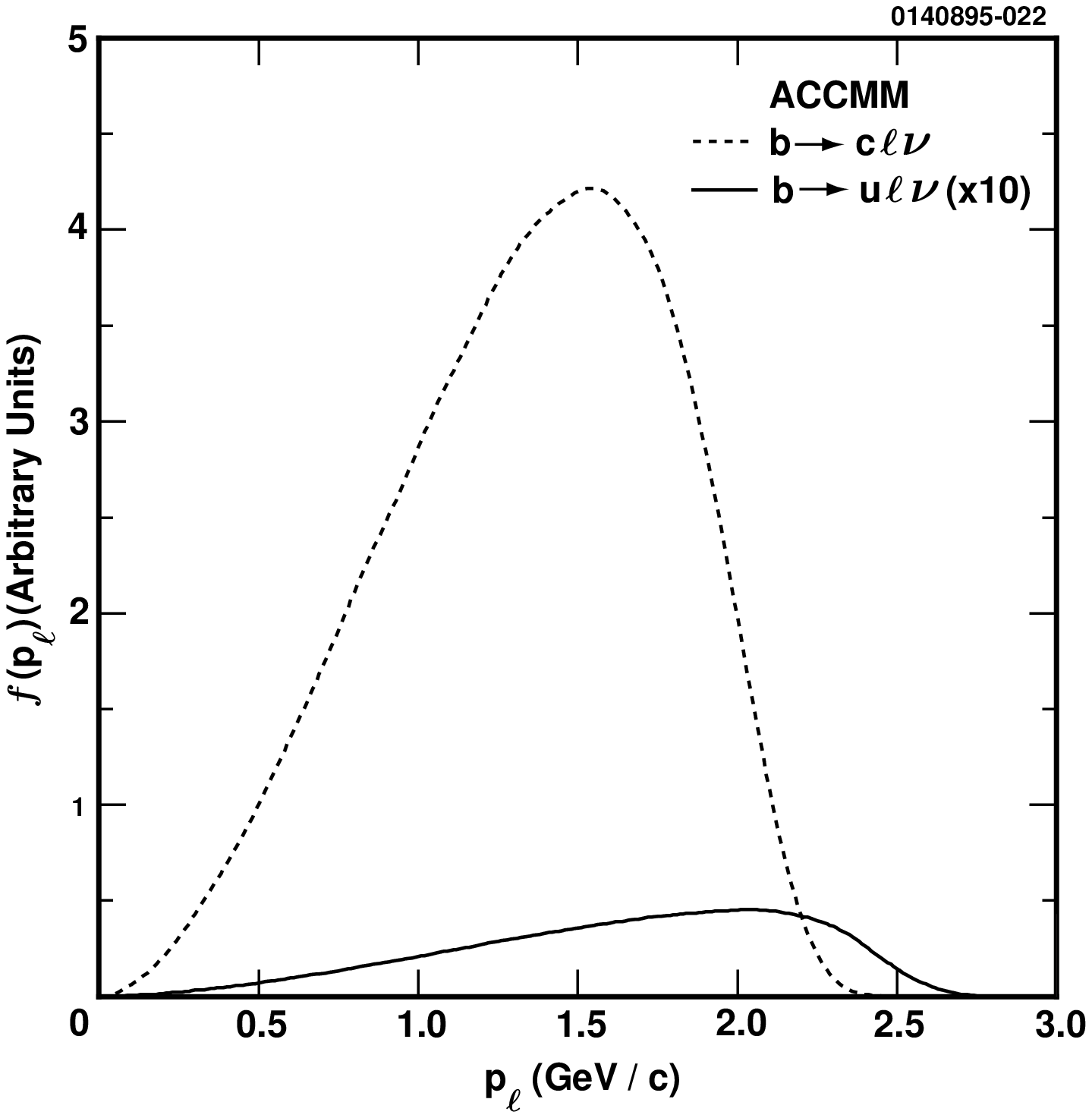}
\includegraphics[height=0.3\textheight]{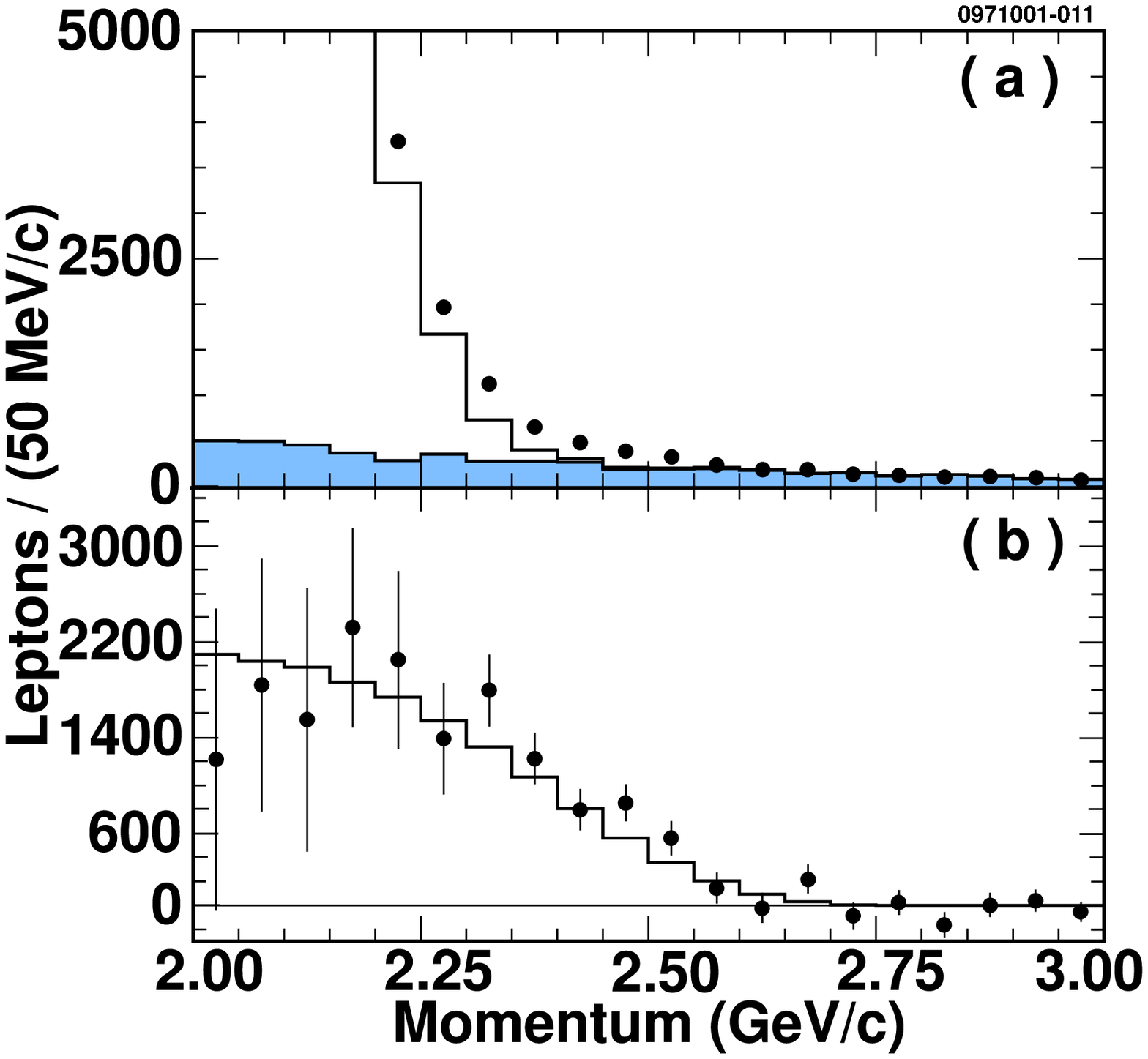}
\label{fig:leptons}
\end{figure}

One might ask, why not do another expansion in 
$\bar{\Lambda}, \lambda_1, \lambda_2$ for the $\bar{B} \to X_u \ell \bar{\nu}$ 
partial width?  One can do this for the fully inclusive 
width; for a discussion of subtleties, see \cite{Uraltsev:1999rr}.  
But in experiments there are very large $b \to c$ backgrounds and 
we can therefore only measure the portion of the rate 
near the lepton momentum endpoint; that is, only above some $p^\ell_{min}$.  
The $b \to X_c \ell\bar{\nu}, X_u\ell\bar{\nu}$ 
spectra are shown in the left half of Figure \ref{fig:leptons}.  
The difficult calculation is the fraction of the leptons above a certain 
momentum cut.  No only do we rely on local duality now, 
but terms of order $1/(M_B - 2 p^\ell_{min})$ can enter, 
spoiling convergence.  
In more physical terms, we require the detailed shape and 
normalization of the spectrum near the endpoint.  

Theory can profitably relate the endpoint $b \to u \ell \bar{\nu}$ rate 
to the observed $b \to s\,\gamma$ spectrum, 
since they are smeared by a common non-perturbative structure function  
\cite{Neubert:1994um,Leibovich:2000ey}, 
up to corrections of order $\Lambda_{QCD}/m_b$ 
(see \cite{Wise:2001} for a review).  
We can extract the structure function from $b \to s\,\gamma$ 
and then use this to predict the fraction of $b \to u \ell \bar{\nu}$ rate 
above the experimental lepton momentum cut. 
There is still an active debate concerning the details of 
the particular methodology we employ \cite{Rothstein:hf}.  

A neural net is used for continuum suppression and 
the signal region in lepton momentum comprises $2.2 < p_\ell < 2.6$ GeV/$c$.  
We have lowered our cut from 2.3 GeV/$c$ to increase the rate.  
The data are shown in the right half of Figure \ref{fig:leptons}.  
We observe good subtraction for $p_l > 2.6$ GeV/$c$, and extract 
$(1874 \pm 123 \pm 326) \, \bar{B} \to X_u \ell \bar{\nu}$ events.   
This yields a partial branching ratio (before radiative corrections) of 
$\Delta {\cal B}_{ub}(2.2 - 2.6\,{\rm GeV}/c) 
   = (2.35 \pm 0.15 \pm 0.45) \times 10^{-4}$.  
Systematics include variations of form factors and heavy charm states 
in Monte-Carlo modeling of $b \to c\ell\bar{\nu}$ backgrounds.  

To extract $|V_{ub}|$, we start with an expression for the inclusive 
rate derived in the upsilon expansion \cite{Hoang:1998hm}: 
\begin{eqnarray}
|V_{ub}| &=& [(3.06 \pm 0.08 \pm 0.08)\times 10^{-3}] \times 
[ ({\cal B}_{ub}/0.001) \cdot (1.6\,{\rm ps}/\tau_B)]^{1/2} 
\nonumber
\end{eqnarray}
The required ${\cal B}_{ub} \equiv B(\bar{B} \to X_u \ell\bar{\nu}$) 
is related to the observed rate $\Delta{\cal B}_{ub}$ 
in momentum window $(p)$ by  
$\Delta {\cal B}_{ub}(p) = F_u(p) \, {\cal B}_{ub}$.  
The $b \to s\,\gamma$ spectrum will provide our prediction for $F_u(p)$, 
which is simply a properly normalized integral of the observed portion 
of the solid curve on the left of Figure \ref{fig:leptons}.  
Using $b \to s\,\gamma$ data with $1.5 < E_\gamma < 2.8$ GeV, 
we fit the shape function \cite{Kagan:1998ym} to various parameterizations.  
We use this to determine \cite{DeFazio:1999sv} 
$F_u(2.2 - 2.6 \,\,{\rm GeV}/c) = 0.138 \pm 0.034 $.  
Consistent results are obtained by the more model-dependent method of 
fitting the spectrum to parameters in the Ali-Greub spectator model 
\cite{Ali:1995bi} and feeding this information into the ACCMM model 
\cite{Altarelli:1982kh,Artuso:1993fb}.  

Our \emph{preliminary} result is:  
\begin{eqnarray}
|V_{ub}| &=& (4.09 \pm 0.14 \pm 0.66) \times 10^{-3}
\nonumber
\end{eqnarray}
This compares favorably with CLEO's exclusive $(\pi/\rho/\omega)\ell\bar{\nu}$ 
analyses \cite{Alexander:1996qu,Behrens:1999vv}:
\begin{eqnarray}
|V_{ub}| &=& (3.25 \pm 0.14^{+0.21}_{-0.29} \pm 0.55) \times 10^{-3}
\nonumber
\end{eqnarray}

\section{Zero-Recoil Point of $\bar{B} \to D^* \ell \bar{\nu}$}

The newer, more inclusive methods above may have reduced model dependence in 
some sense, but quark-hadron duality is always involved.  We now turn to 
a more traditional exclusive extraction of $|V_{cb}|$ from 
$\bar{B} \to D^* \ell \bar{\nu}$ decays.  

This analysis measures the absolute rate as a function of $q^2$.  
Both $D^{*+} \ell \bar{\nu}$ and $D^{*0} \ell \bar{\nu}$ modes 
are reconstructed, 
with $0.8 < p_e < 2.4$ GeV/$c$ and $1.4 < p_\mu < 2.4$ GeV/$c$.  
One usually replaces $q^2$ with $w = \vec{v}_B \cdot \vec{v}_D$, 
the product of $B$ and $D^*$ meson four-velocities.  
This new variable is just a particular linear transform, $w = a - b q^2$.  
HQET simplifies the analysis by relating the three form factors present 
to one universal Isgur-Wise function.  
In fact for $\bar{B} \to D^* \ell \bar{\nu}$, the point $w = 1$ 
(corresponding to maximum $q^2$) 
has no ${\cal O}(1/M)$ corrections \cite{Luke:1990eg}.  
Physically, this occurs when $\vec{v}_B = \vec{v}_D$, 
that is at \emph{zero-recoil} where the $D^*$ 
is at rest relative to the $B$.  
This is the favorable place for a precise extraction of $|V_{cb}|$.  

\begin{figure}
\caption{Left: Signal and backgrounds for $\bar{B} \to D^* \ell \bar{\nu}$ 
         in the bin $1.10 < w < 1.15$ for both $D^{*+}$ and $D^{*0}$ 
         modes.  Note that backgrounds can be constrained from rates 
         in the non-physical region of the $\cos \theta_{B - D^*\ell}$.  
         Combinatoric background refers to fake $D^*$; 
         (un)correlated background refers to cases where the 
         real $D^*$ and $\ell$ are (not) from the same $B$ decay.  
         Right: The top two panels show the fit yields in each $w$ bin 
         for each $D^*$ mode, while the lower panel shows the 
         extracted form factor with the final fit overlaid.  }
\includegraphics[height=0.45\textheight]{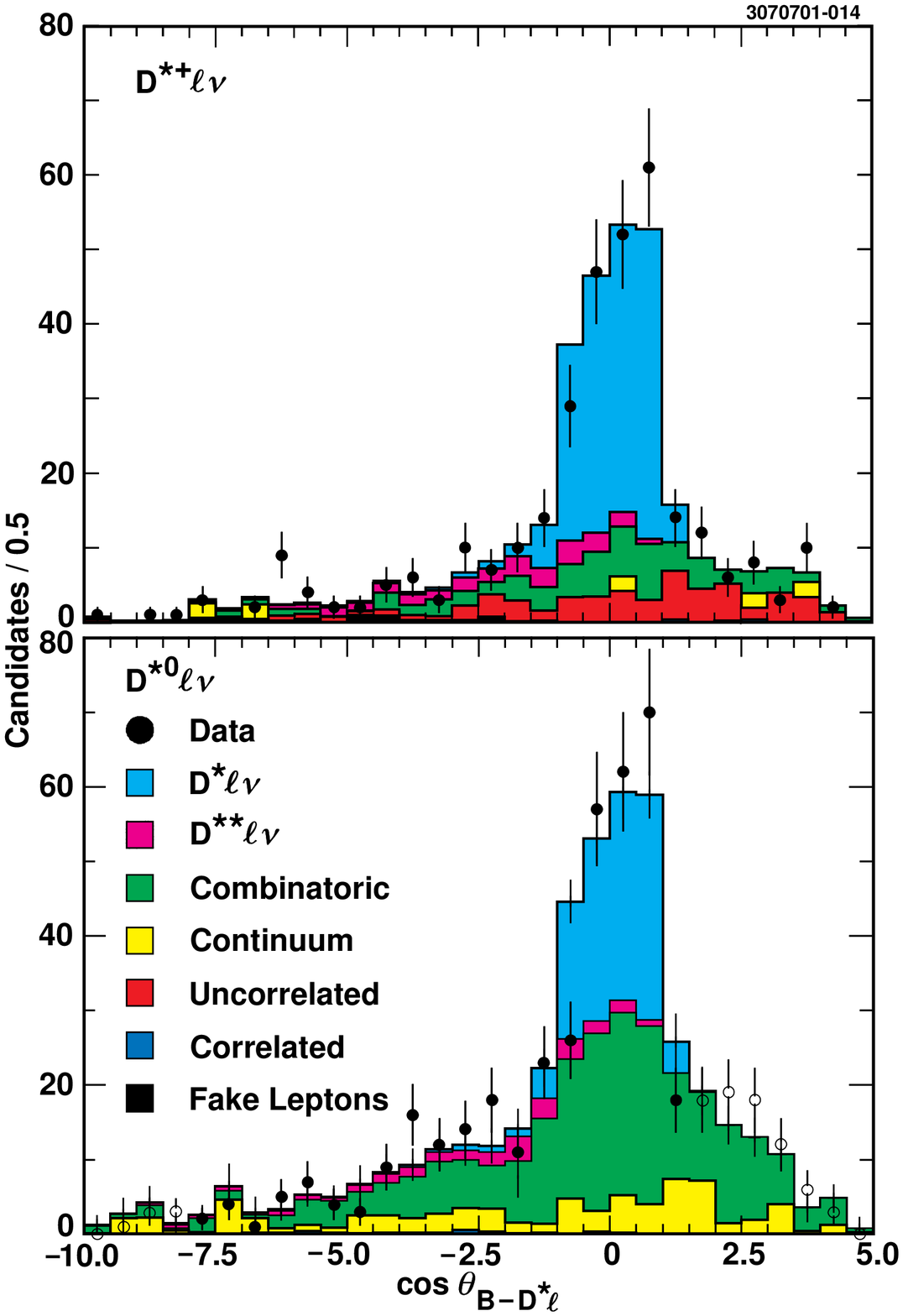}
\includegraphics[height=0.45\textheight]{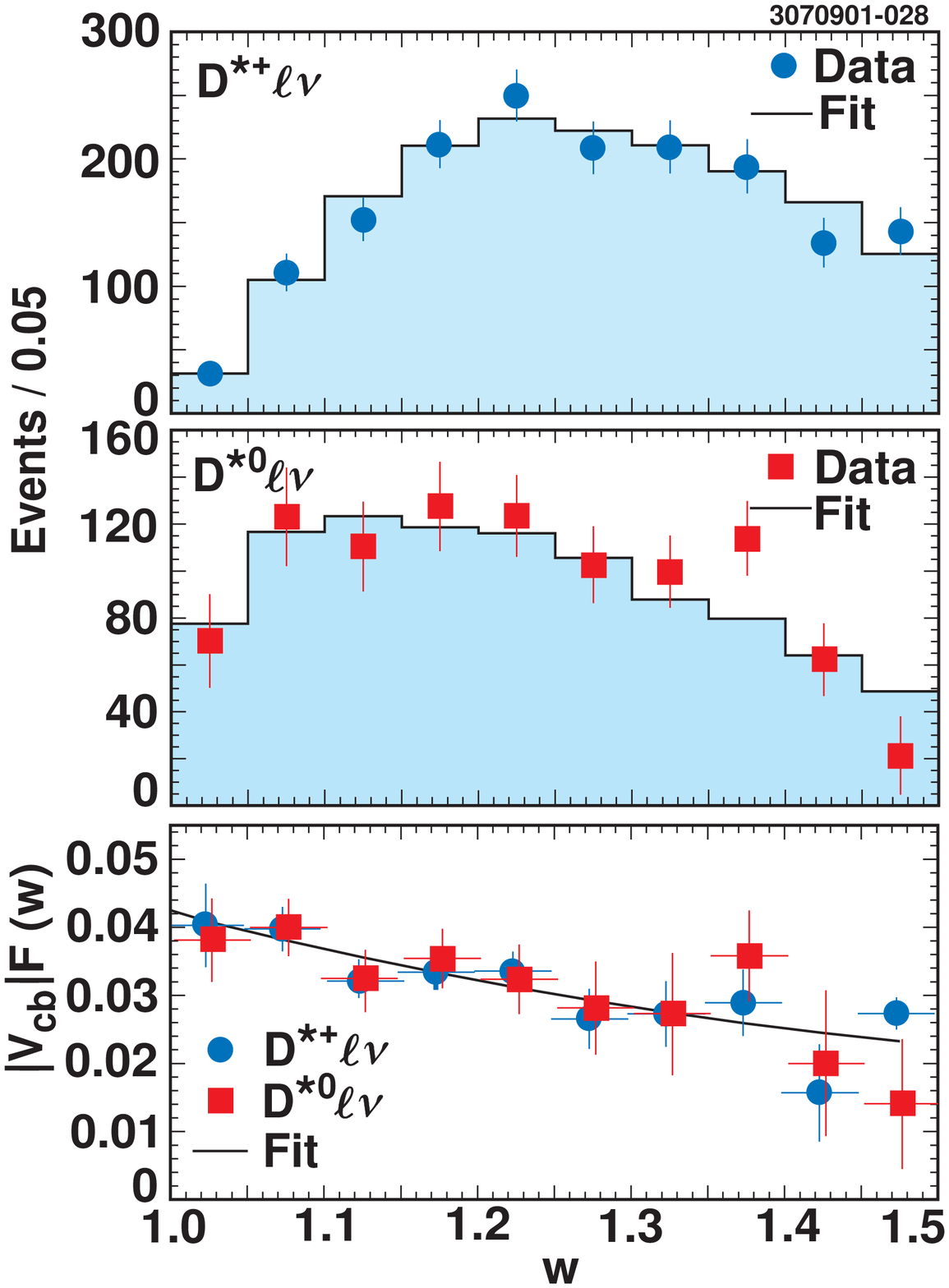}
\label{fig:dlnu_new}
\end{figure}

Background discrimination is accomplished by 
examining the angle, $\theta_{B - D^*\ell}$, 
between the $B$ and the $D^*\ell$ system.  
This variable is similar to the familiar missing-mass.  
Since $\cos \theta_{B - D^*\ell}$ is calculated from four vectors it 
can be unphysical and its shape can be helpful in disentangling the 
many types of background.  
Determining the rate for each bin in $w$ requires a detailed fit 
like the examples on the left of Figure \ref{fig:dlnu_new}.  
The right half of the figure shows both the fit rate and, after efficiency 
and kinematic factors, the extracted form-factor as a function of $w$.  
Inclusion of the $D^{*0}$ mode adds significantly more efficiency at 
zero recoil as compared to using $D^{*+}$ alone.  

We determine the \emph{preliminary} branching ratios: 
\begin{eqnarray}
{\cal B}(\bar{B}^0 \to D^{*+} \ell \bar{\nu}) &=& (5.82 \pm 0.19 \pm 0.37)\%
\nonumber \\
{\cal B}(      B^- \to D^{*0} \ell \bar{\nu}) &=& (6.21 \pm 0.20 \pm 0.40)\%
\nonumber
\end{eqnarray}
The intercept at zero-recoil and slope of the HQET form-factor 
(with the curvature related to the slope by dispersion relations 
\cite{Boyd:1997kz,Caprini:1998mu}) are \emph{preliminarily} determined as: 
\begin{eqnarray}
   F(1) |V_{cb}| &=& (4.22 \pm 0.13 \pm 0.18) \times 10^{-2}
\nonumber \\
   \rho^2 &=& 1.61 \pm 0.09 \pm 0.21
\nonumber
\end{eqnarray}

Significant systematics include efficiency (especially for slow pions), 
the intermediate branching ratios, backgrounds, 
and (especially for $\rho^2$) the ratios of the three $D^*$ form factors 
$R_1$ and $R_2$ \cite{Duboscq:1996mv}.  
Using $F(1) = 0.913 \pm 0.042$  \cite{Harrison:1998yr}, we extract:
\begin{eqnarray}
 |V_{cb}| &=& (4.62 \pm 0.14 \pm 0.20 \pm 0.21) \times 10^{-2}
\nonumber
\end{eqnarray}
where the errors are statistical, systematic, and theoretical, 
for a net 7\% precision.  

\begin{figure}
\caption{A graphical compilation of the present result along with 
        published results from LEP experiments \cite{LHFWG:2001}.  
        'OPAL inc' denotes an analysis using partial reconstruction 
        of $D^{*+} \ell \bar{\nu}$.  
        Ellipses indicate $\Delta \chi^2 = 1$, including systematic errors.}
\includegraphics[height=0.25\textheight]{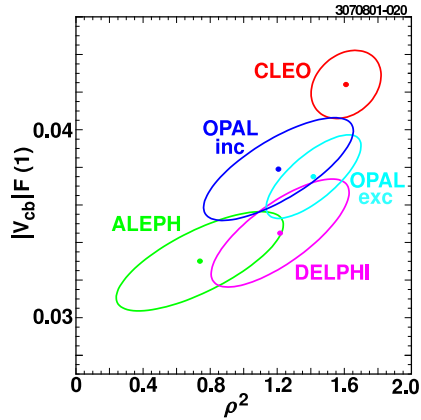}
\label{fig:dlnusummary}
\end{figure}

A comparison of recent results for the slope and intercept of the 
$\bar{B} \to D^* \ell \bar{\nu}$ form factor is shown 
in Figure \ref{fig:dlnusummary}.  
The differing shape of the CLEO ellipse is due to an interaction of 
the lepton momentum cut with variations of the form-factor ratios 
$R_1$ and $R_2$ within their errors.  
There is a disagreement at about the two sigma level.  
One difference in technique involves the $D^* X \ell \bar{\nu}$ background;  
CLEO includes these in the $\cos \theta_{B - D^*\ell}$ fits to the data, 
while LEP analyses use a model constrained to other LEP results on 
$\bar{B} \to D^* X \ell \bar{\nu}$.  

\section{The Future}

There are several related analyses in progress at CLEO.  
These include 
$\bar{B} \to X_c \ell \bar{\nu}$ lepton spectrum moments ($<E_\ell>$) 
which will provide another band in the $\bar{\Lambda} - \lambda_1$ plane.  
Both low-background tagged (di-lepton) and higher statistics untagged 
analyses are being pursued.  
We also have more statistics for the exclusive 
$\bar{B} \to D^* \ell \bar{\nu}$ analysis.  

Branching ratios and form-factor investigations 
for the $\bar{B} \to (\pi/\rho/\omega) \ell \bar{\nu}$ modes
 used for $|V_{ub}|$ are underway using neutrino reconstruction.  
We will further address inclusive measures of $|V_{ub}|$ with 
inclusive leptons, but now making full use of kinematics.   
Instead of singling out the lepton momentum, one can use quantities 
such as $q^2$ and the recoil mass \cite{Falk:1997gj}.  
We hope to accept a larger portion of rate while controlling background, 
thus reducing uncertainty on fraction of the 
$b \to u \ell \bar{\nu}$ rate observed.  

Results for $|V_{cb}|$ from $D^{*+} e^+ \bar{\nu}$ \cite{Abe:2001dstar} 
and from ${\cal B}(\bar{B} \to X e^- \bar{\nu})$ \cite{Abe:2001incl} 
have recently been presented by Belle.  
Future results from both Belle and BaBar will be of great interest.  

\section{Conclusion}

We have measured the photon spectrum from $b \to s\,\gamma$ decays 
and the hadronic mass moments from $\bar{B} \to X_c \ell \bar{\nu}$.  
Using HQET and ${\cal B}(\bar{B} \to X_c \ell \bar{\nu})$, 
we extract $|V_{cb}|$ with more controlled theoretical systematics, 
but still subject to duality issues.  
New techniques relating studies of the lepton endpoint 
using a structure function constrained 
to $b \to s\,\gamma$ photon spectrum allow us to extract $|V_{ub}|$.  
Lastly, we update our $|V_{cb}|$ result using the $\bar{B} \to D^* \ell \bar{\nu}$ 
rate extrapolated to zero recoil.  
Ongoing analyses will extend all of this work further.  
Our emphasis is on using a variety of techniques, with differing systematics.  

\begin{theacknowledgments}
It is a pleasure to thank my CESR and CLEO colleagues for their 
sustained efforts.  Our research is supported by the NSF and DOE.  
Current information and results may be found at 
\url{http://w4.lns.cornell.edu/public/CLEO/}

\end{theacknowledgments}

\hyphenation{Post-Script Sprin-ger}

\end{document}